\documentclass[aps,prb,twocolumn,showpacs,amsmath,amssymb,superscriptaddress]{revtex4}
\usepackage{dcolumn}
\usepackage{bm}
\usepackage{graphicx}
\usepackage{times}
\usepackage{epstopdf}
\begin{document}

\makeatletter
\newcommand*{\balancecolsandclearpage}{%
  \close@column@grid
  \clearpage
  \twocolumngrid
}
\makeatother

\title{Odd-frequency superconducting pairing in topological insulators}
\author{Annica M. Black-Schaffer}
 \affiliation{Department of Physics and Astronomy, Uppsala University, Box 516, S-751 20 Uppsala, Sweden}
 \author{Alexander V. Balatsky}
\affiliation{Theoretical Division and Center for Integrated Nanotechnologies, Los Alamos National Laboratory, Los Alamos, New Mexico 87545, USA}
\affiliation{Nordic Institute of Theoretical Physics (NORDITA), Roslagstullsbacken 23, S-106 91 Stockholm, Sweden}
\date{\today}

\begin{abstract}
We discuss the appearance of odd-frequency spin-triplet $s$-wave superconductivity, first proposed by Berezinskii [{\it JETP} {\bf 20}, 287 (1974)], on the surface of a topological insulator proximity coupled to a conventional spin-singlet $s$-wave superconductor. Using both analytical and numerical methods we show that this disorder robust odd-frequency state is present whenever there is an in-surface gradient in the proximity induced gap, including superconductor-normal state (SN) junctions. The time-independent order parameter for the odd-frequency superconductor is proportional to the in-surface gap gradient. The induced odd-frequency component does not produce any low-energy states.
\end{abstract}
\maketitle

%
\section{Introduction}
Topological insulators (TIs) are a new class of materials \cite{Hasan10, Qi11} with an insulating bulk but with a conducting surface state. The surface state has its spin locked to the momentum in a Dirac-like energy spectrum.
Superconducting TI surfaces have received a lot of attention recently, \cite{Beenakker11, Alicea12} since it was predicted that Majorana modes appear in e.g.~superconducting vortex cores and at superconductor-ferromagnet (SF) interfaces.\cite{Fu08, Fu09} Experimentally, both superconducting transport\cite{Sacepe11} as well as the Josephson effect\cite{Veldhorst12} have already been demonstrated in TIs proximity-coupled to conventional superconductors.
Despite this large interest, relatively little attention has been paid to the superconducting state itself.
In addition to the standard proximity effect, one could expect the spin-orbit coupling in TIs to lead to significant modifications and produce novel superconducting states that are not easily accessible in conventional superconductors. For example, it has already been demonstrated that an effective $p$-wave pairing is induced when the TI is proximity-coupled to a conventional $s$-wave superconductor.\cite{Fu08,Stanescu10, Black-Schaffer11QSHI}

Quite generally, the superconducting pair amplitude, being the wave function of the Cooper pairs, needs to obey Fermi-Dirac statistics. This leads to the traditional classification into even-parity ($s$, $d$, ...) spin-singlet and odd-parity ($p$, $f$, ...) spin-triplet pairing. It has also been shown that the pair amplitude can be odd in frequency.\cite{Berezinskii74, Balatsky92}
Odd-frequency spin-triplet $s$-wave pairing has been found in spin-singlet $s$-wave SF junctions due to spin-rotational symmetry breaking and it explains the long-range proximity effect in these junctions.\cite{Bergeret01, Bergeret05} Very recently, the same magnetic field induced odd-frequency pairing has also been found in TIs.\cite{Yokoyama12}
In superconductor-normal metal (SN) junctions translational symmetry breaking instead generates odd-frequency spin-singlet odd-parity components \cite{Tanaka07JJ, Tanaka07PRB}. However, the odd-parity limits the odd-frequency pairing to ballistic junctions.\cite{Tanaka07}

In this article we show with a simple analysis that the effective spin-orbit coupling ${\bf k} \cdot \sigma$ on the TI surface immediately induces odd-frequency spin-triplet $s$-wave correlations, even in the absence of a magnetic field. The $s$-wave nature of the odd-frequency component makes it robust against disorder, in sharp contrast to normal SN junctions.
The odd-frequency correlations appear whenever there is an in-surface gradient in the proximity-induced spin-singlet $s$-wave pairing, with the odd-frequency order parameter directly proportional to the in-surface gradient.
We numerically calculate the odd-frequency response in several superconducting two-dimensional (2D) TI systems, including SN, SS' junctions, and in the presence of surface supercurrents. 
These results point to an important missing component in the discussion on the role of proximity induced superconductivity in TIs and the odd-frequency component ought to be included in the study of low energy states and Majorana fermions in TIs.
We also discuss experimental consequences of the odd-frequency pairing. We find that the analytic  $~1/\omega$ form of the odd-frequency response does not result in low energy states, which previously has been intimately linked to the appearance of odd-frequency components,\cite{Tanaka12, Asano12} but the predicted gapped spectrum could allow detection of the odd-frequency component with local tunneling probes. The spin-triplet pairing will further produce a finite Knight shift in nuclear magnetic resonance (NMR) or muon spin-rotation measurements. 

%
\section{Analytic derivation}
We start with an analytic calculation that illustrates the appearance of an odd-frequency component. The Hamiltonian that describes the contact between a TI and a conventional $s$-wave spin-singlet superconductor (SC), see Fig \ref{fig:SN}(a), can be written as $ H = H_{\rm TI} + H_{\rm SC}+ H_{T}$:
\begin{align}
\label{eq:Htot}
H_{\rm TI} & = \sum_{\bf{k}, \alpha, \beta} c_{\alpha, \bf{k}}^{\dag} {\bf k} \cdot \sigma_{\alpha \beta} c_{\beta, {\bf{k}}} \\ \nonumber
H_{\rm SC} & = \sum_{\bf k, \alpha, \beta} \varepsilon({\bf k}) d_{\alpha, \bf k}^{\dag}d_{\alpha, \bf k} + \sum_{i, \alpha, \beta}\Delta(i)_{ \alpha \beta} d^\dagger_{\alpha, i}d^\dagger_{\beta, i} + {\rm H.c.}\\ \nonumber
H_{T} & = \sum_{\alpha} T_i c_{\alpha, i}^{\dag}d_{\alpha, i} + {\rm H.c.}.
\end{align}
$H_{\rm TI}$ is the Hamiltonian describing the TI surface state at $y < 0$, $H_{\rm SC}$ is the Hamiltonian for the SC electronic states at $y > 0$, and $H_{T}$ describes the tunneling between the SC and TI.
We use the low energy approximation around the Dirac point for the TI dispersion. $H_{\rm SC}$ is of the standard form, where we explicitly allow for position dependence through the site index $i$, where then $\hat{\Delta} = \Delta(i)_{ \alpha \beta} =  \epsilon_{\alpha \beta} \Delta(i)$ is the matrix in spin space, which we choose to be spin-singlet. We assume that the kinetic energy $\varepsilon({\bf k})$ in the SC is a simple band and that the tunneling matrix element $T_i$ is nonzero only for neighboring sites at the TI-SC interface.
We further assume that $\Delta(i)$ is dependent on the in-surface $x$-coordinate and approximate it by a linear slope $\Delta(i) = \Delta_0 + (i a) {\frac{\partial \Delta} {\partial x}}|_0$ where $a$ is the unit cell size.
Superconducting correlations in the TI will be induced by pairs tunneling into the TI from the superconducting side. Therefore, any induced pair amplitude in the TI will be proportional to $\Delta(i)$ and its derivatives at the interface. To derive the induced pair component on the TI surface, we evaluate the anomalous Green's function $F_{{\rm TI}, \alpha \beta}({\bf k}, {\bf k'}) = -\imath \langle T_{\tau}c_{\alpha}(\tau, {\bf k})c_{\beta}(0, {\bf k'})\rangle$. Using standard methods we find it to be proportional to
\begin{align}
\label{eq:Ganom}
\hat{F}_{\rm TI}(\omega_n|i, i) = -|T|^2 \sum_{j,l} \hat{G}^0(\omega_n|i,j) \hat{F}(j,l)\hat{G}^0(\omega_n|l,i),
\end{align}
where $\hat{G}^0(\omega|{\bf k}) = ({\bf k} \cdot \sigma - i \omega_n)/ ({\bf k}^2 + \omega_n^2)$ is the Matsubara Green's function for the electrons in the TI, $\hat{F}(\omega_n, {\bf k}) = \hat{\Delta}(i)[\omega_n^2 + \varepsilon^2({\bf k}) + \hat{\Delta}(i)^2]^{-1}$ is the anomalous Green's function for a conventional superconductor, where we assume slow variations of $\hat{\Delta}(i)$. By going to momentum space and using the $k$-space expansion $\Delta({\bf k}) = \Delta_0 \delta_{{\bf k}, 0} + {\frac{\partial \Delta} {\partial x}}|_0 \imath \partial_{{\bf k}_x}$, we can rewrite the nontrivial part of the induced pair amplitude on the surface of the TI as
 \begin{align}
 \label{eq:Ganom2}
\hat{F}_{\rm TI}(\omega_n|i = 0) & \! = \! \! \sum_{\bf k} \frac{-\imath |T|^2 \partial_x \hat{\Delta}|_0 \hat{G}^0(\omega_n|{\bf k}) \partial_{{\bf k}_x}\hat{G}^0(\omega_n|\!-\! {\bf k})}{ 2[\omega_n^2 \!+ \! \varepsilon^2({\bf k})\! + \! \Delta^2(0)]}  \nonumber \\
& = \sum_{{\bf k}}  \frac{ |T|^2 \omega_n \hat{\sigma} \partial_x \hat{\Delta}|_0} {2[\omega_n^2 + \varepsilon({\bf k})^2 + \Delta^2(0)](\omega_n^2 + {\bf k}^2)^2}.
\end{align}
This result indicates that the Josephson tunneling into the TI will immediately induce an {\it odd-frequency spin-triplet even-momentum} superconducting component.\cite{Berezinskii74}
It is the characteristic spin-momentum locking in the Dirac surface spectrum that induces spin-triplet odd-frequency pairing in the presence of translational symmetry breaking. The $s$-wave nature of this pairing guarantees robustness against disorder. The situation here is different from the normal metal case where only odd-frequency spin-singlet odd-parity correlations are induced by translational symmetry breaking.\cite{Tanaka07JJ, Tanaka07PRB}
To evaluate the on-site odd-frequency component we will assume that  $E_F$ of the conventional supercoductor is the dominant scale and we ignore higher order terms in $1/E_F^2$. The local on-site amplitude on the TI surface is given by an integral over the momenta and is proportional to $\hat{F}_{\rm TI}(\omega_n| i=0) \sim  |T|^2 \omega_n \sigma^z \partial_x {\Delta}|_0/ (E^2_F |\omega_n|^2) $. Interestingly, $1/w$ dependence has also been reported in heavy fermion compounds.\cite{Coleman93}
The particular form of the gap function allows us to introduce an order parameter, i.e.~the inherent parameter that is constant at equal time in the odd-frequency superconductor, see e.g.~\onlinecite{Dahal09}. The odd-frequency order parameter in our case is proportional to
\begin{eqnarray}
\partial_{\tau} \hat{F}_{\rm TI}(\tau| i)|_{0} \sim \sum_n  |T|^2\sigma^z \frac{\omega^2_n}{|\omega_n|^2}  \frac{\partial {\Delta}}{\partial{x}} \sim \frac{\partial{\Delta}}{\partial{x}}.
\label{eq:OP}
\end{eqnarray}
This proportionality to the in-plane gradient of the $s$-wave gap can be tested and we indeed find that  $\partial_{\tau} \hat{F}_{\rm TI}(\tau| i)|_{0}$ is tracking the spatial gradient of the gap, see Figs.~\ref{fig:SN}(f) and \ref{fig:curr}(b).

One of the observable consequences of odd-frequency superconductivity is usually the appearance of sub-gap states,\cite{Tanaka07JJ,Tanaka07PRB, Yokoyama07, Tanaka12, Asano12} or even a low-energy continuum associated with a gapless nature of the state.\cite{Balatsky92, Dahal09}
However, we find here that the particular structure of the odd-frequency gap $\sim 1/|\omega_n|$ results in no intragap states at the lowest energies and thus this odd-frequency gap state is fully gapped. Indeed, after analytic continuation from the Matsubara axis, the local density of states (LDOS) is proportional to $N(E) \sim {\rm Re}[E/\sqrt{E^2 - \Delta(E)^2}] \sim {\rm Re}[1- C^2/E^2]^{-1}$, which vanishes at energies below the minigap induced by the tunneling $C \sim |T|^2 \partial {\Delta}/\partial{x}$.  One has to keep in mind that these features will occur in addition to the LDOS features introduced by the induced conventional BCS pairing.

%
\section{Numerical results}
The analytical results in Eqs.~(\ref{eq:Ganom2}-\ref{eq:OP}) are derived for a 3D TI, but are equally valid for a 2D TI. To complement these results we explicitly calculate the odd-frequency spin-triplet superconducting correlations in the Kane-Mele 2D TI:\cite{Kane05b}
\begin{align}
\label{eq:H0}
H_{\rm KM}  =   -t \sum_{\langle i,j\rangle} c_{i}^\dagger c_{j} + \! \mu \sum_{i} c_{i}^\dagger c_{i} +
i \lambda \sum_{\langle \langle i,j\rangle \rangle} \nu_{ij} c_{i}^\dagger \sigma^zc_{j},
\end{align}
where $c_{i}^\dagger$ is now the fermion creation operator on site $i$ in the honeycomb lattice with the spin-index suppressed. Furthermore, $\langle i,j \rangle$ and $\langle \langle i,j \rangle \rangle$ denote nearest neighbors and next-nearest neighbors respectively, $t$ is the nearest neighbor hopping amplitude, $\mu$ the chemical potential, $\lambda$ the spin-orbit coupling, and $\nu_{ij} = +1$ $(-1)$ if the electron makes a left (right) turn to get to the second bond. We set the energy and length scales by fixing $t = 1$ and $a = 1$, respectively and choose $\lambda = 0.3$, which gives a bulk band gap of $1$.
%
The influence of the SC can be described by an effective on-site attractive potential $U_i$ acting at the TI edge when it is in proximity to a SC:\cite{Black-Schaffer11QSHI,backactionnote}
\begin{align}
\label{eq:HSC}
H_{\Delta}  =   -\sum_i U_i c_{i\downarrow} c_{i\uparrow} c^\dagger_{i\uparrow} c^\dagger_{i\downarrow}.
\end{align}
We solve $H = H_{\rm KM} + H_\Delta$ self-consistently for the spin-singlet $s$-wave mean-field order parameter $\Delta(i) = -U_i \langle c_{i\downarrow} c_{i\uparrow}\rangle$.
%
We further introduce the odd-frequency spin-triplet $s$-wave pairing correlations:\cite{Halterman07}
\begin{align}
\label{eq:Ft0}
F_t^0(\tau | i) & = \langle c_{i \uparrow}(\tau)c_{i \downarrow}(0) +  c_{i \downarrow}(\tau)c_{i \uparrow}(0)\rangle/2, \\
\label{eq:Ft1}
F_t^1(\tau | i) & = \langle c_{i\uparrow}(\tau)c_{i \uparrow}(0) -  c_{i \downarrow}(\tau)c_{i \downarrow}(0)\rangle/2
\end{align}
where $\tau$ is the time coordinate (at zero temperature). Eqs.~(\ref{eq:Ft0}-\ref{eq:Ft1}) contain all space and time information of the odd-frequency spin-triplet response.
As required by the Pauli principle, $F_t(\tau = 0)$ always vanishes for a self-consistent solution of $\Delta$, whereas the time derivative at equal times $\partial_\tau F_t \left|_0 \right.$ defines the odd-frequency order parameter, in analogy with Eq.~(\ref{eq:OP}). Also, since $H_{\rm KM}$ commutes with $\sigma^z$, the $m = 1$ spin-triplet projection $F_t^{1}$ is identically zero.
We also define a time-dependent quantity for $s$-wave spin-singlet pairing: $F_s(\tau | i) = (\langle c_{i \downarrow}(\tau)c_{i \uparrow}(0) - c_{i \uparrow}(\tau)c_{i \downarrow}(0)\rangle)/2$, which is related to the order parameter through $\Delta_i = -U_i F_s(\tau = 0 | i)$.

\subsection{SN junction}
\begin{figure}[htb]
\includegraphics[scale = 0.9]{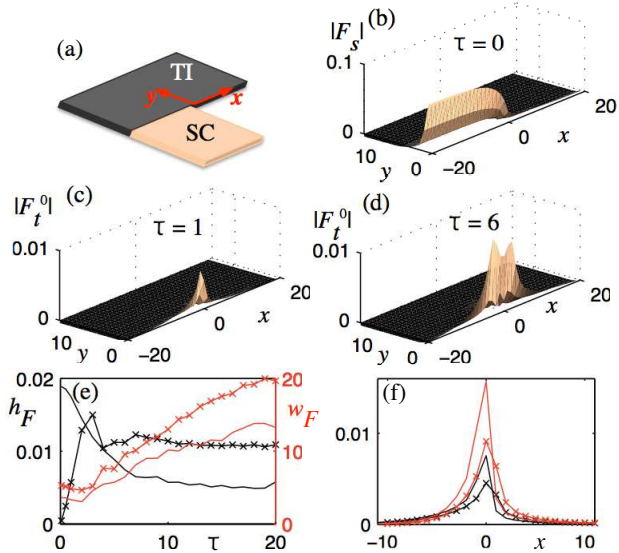}
\caption{\label{fig:SN} (Color online) (a) SN junction along the edge of a 2D TI. (b) Even-frequency pairing $|F_s(0)|$, (c) odd-frequency pairing $|F_t^0(1)|$, and (d) $|F_t^0(6)|$ in a 2D TI SN junction with $\mu_{\rm S} = 0.3$, $\mu_{\rm N} = 0$, and $U = 2.2$, giving $\Delta_{\rm S} = 0.16$. (e) Maximum height $h_F$ (black, left axis) and average width $w_F$ (red, right axis) as function of $\tau$ for $|F_t^0|$ (crosses) and $|\partial_x F_s|$ (line). (f) Magnitude of odd-frequency order parameter $\partial_\tau F_t^0\left|_0 \right.$ (crosses) and $\frac{\partial \Delta}{\partial x}$ scaled by a factor of 0.4 (line) for $\Delta_{\rm S} = 0.16$ (black) and $\Delta_{\rm S} = 0.30$ (red).
 }
\end{figure}

From Eq.~(\ref{eq:Ganom2}) we know that non-zero odd-frequency spin-triplet correlations require a gradient in the superconducting order parameter along the TI edge, i.e.~$\frac{\partial \Delta}{\partial x}$ needs to be finite. This is the case e.g.~at any step edge in a TI proximity coupled to a SC, but a more striking example is a SN junction along the TI edge, as schematically pictured in Fig.~\ref{fig:SN}(a). In the S region of the TI, we apply a constant $U$, and, since the SC provides an ample source of charge, we also set $\mu_{\rm S}>\mu_{\rm N}$.
%
In Fig.~\ref{fig:SN}(b) we plot the magnitude of the self-consistent spin-singlet pairing amplitude $F_s(\tau = 0)$, which shows proximity-induced superconducting pairing in the N region, as well as an inverse proximity effect (depletion of Cooper pairs) on the S side of the junction. 
The odd-frequency response is shown in Figs.~\ref{fig:SN}(c-e), with $F_t$ only non-zero in the SN interface region where the gradient of $\Delta$ is finite. For small $\tau$, $F_t$ is sharply peaked at the interface with the peak height $h_F$ rapidly increasing with $\tau$, as seen in Fig.~\ref{fig:SN}(e). For $\tau \gtrsim 4$ (small frequencies $\omega_n$) the height is approximately constant, but the peak instead becomes broader, with the average width $w_F$, defined as the ratio of the total weight of the peak to its height $h_F$, increasing roughly linearly with $\tau$.
In Fig.~\ref{fig:SN}(e) we also plot the height and width of the gradient of the spin-singlet response $\partial_x F_s$ for a direct comparison. For large $\tau$, $F_t$ is directly related to $\partial_x F_s$, supporting the analytic result in Eq.~(\ref{eq:Ganom2}), whereas at small $\tau$ the direct relation breaks down for the peak height.
In Fig.~\ref{fig:SN}(f) we finally compare the odd-frequency order parameter $\partial_\tau F_t\left|_0\right.$ with $\frac{\partial \Delta}{\partial x}$. As predicted in Eq.~(\ref{eq:OP}), we find these two quantities to be directly proportional.

The odd-frequency pairing in a SN TI junction is strikingly different from that of a conventional SN junction. The odd-frequency response in the TI case has even-parity spin-triplet symmetry, which is robust against impurity scattering and can thus survive even in the diffusive regime,\cite{Tanaka07} in contrast to the odd-parity spin-singlet symmetry in a regular SN junction.\cite{Tanaka07JJ, Tanaka07PRB}
Related to this disorder robustness, we also find that $F_t$ is insensitive to any Fermi level mismatch at the SN interface, created by using different chemical potentials in N and S. In a normal SN junction, the odd-frequency pairing quickly disappears when the transparency of the interface is reduced.\cite{Tanaka07PRB}
Finally, to investigate the influence of the odd-frequency pairing on the low-energy spectrum, we study SS' junctions, where the two S regions have different pair potentials $U$. We find no evidence for a reduced gap or intragap states in both sharp and extended interface junctions.\cite{SMnote} This expands our analytical low-energy result to exclude any intragap states, and is again in sharp contrast to conventional SN and SF junctions.\cite{Tanaka07JJ,Tanaka07PRB, Yokoyama07, Tanaka12, Asano12}

\subsection{Supercurrent}
Odd-frequency correlations can also be generated in a homogenous TI-SC system if a supercurrent $I$ is applied in-surface, since the current is proportional to the gradient of the phase of the superconducting order parameter. We model such a  system by setting $\Delta = |\Delta| e^{\imath k x}$, with $k$ being the (fixed) phase winding which is proportional to the current, and solve self-consistently for $\Delta$.
We find that $F_s(\tau) = C_s(\tau) e^{\imath (k x + \theta_s(\tau))}$ and $F_t = C_t(\tau) e^{\imath (k x + \theta_t(\tau))}$, i.e.~both $F_s$ and $F_t$ have the same phase winding $k$ for all $\tau$, but the amplitudes $C$ and phase off-sets $\theta$ are dependent on $\tau$.
\begin{figure}[htb]
\includegraphics[scale = 0.9]{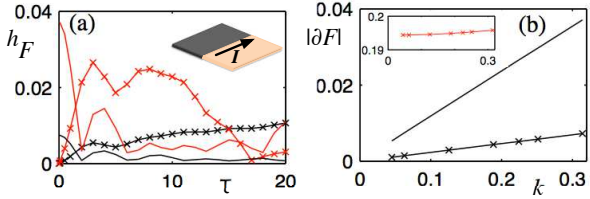}
\caption{\label{fig:curr} (Color online) (a) Magnitude (height) of the odd-frequency pairing $F_t^0$ (crosses) and $\partial_x F_s$ (line) as function of $\tau$ for phase windings $k = \pi/50$ (black) and $\pi/10$ (red) along the TI edge. Inset shows schematically a supercurrent $I$ along the edge of a 2D TI. (b) Magnitude of the odd-frequency order parameter $\partial_\tau F_t^0|_0$ (crosses) and $\frac{\partial \Delta}{\partial x}$ (line) as function of $k$. Inset shows their ratio. Here $U = 2.5$, $\mu = 0.3$, giving $\Delta_{\rm S} = 0.30$.
 }
\end{figure}
In Fig.~\ref{fig:curr}(a) we plot the (position-independent) magnitudes of $F_t$ and $\partial_x F_s$ as function of $\tau$ for two different values of $k$. Similar to the SN junction, $F_t$ increases rapidly for small $\tau$. For small currents $F_t$ continues to increase even for large $\tau$, even though less steeply. This is in contrast to $\partial_x F_s$ which, apart from small oscillations, decreases with increasing time. For larger currents we see a sharp downturn in $F_t$ at intermediate time values. Before this downturn we find that $F_t$ is directly proportional to $k$, as expected from its relation to $\partial_x F_s$ in Eq.~(\ref{eq:Ganom2}). However, for times past the downturn, $F_t$ instead decreases with increasing $k$. The phase off-set parameters $\theta_s$ and $\theta_t$ also tracks each other (with the expected $\pi/2$ shift) before the downturn in $F_t$, but past this time the correlation between them is lost. Thus, we find that $F_t$ and $\partial_x F_s$ are only tracking each other in a finite $\tau$-range, which set by the current.
In Fig.~\ref{fig:curr}(b) we focus on the behavior at $\tau =0$. Both the odd-frequency order parameter $\partial_\tau F_t^0|_0$ and $\frac{\partial \Delta}{\partial x}$ are linear in $k$, with their ratio being a constant for all currents.
This shows that the odd-frequency order parameter is always directly proportional to the current (phase winding), in agreement with Eq.~(\ref{eq:OP}), and thus applying a supercurrent offers an experimentally easy way to tune the strength of the odd-frequency pairing.

\subsection{Rashba spin-orbit coupling}
So far we have only discussed $m = 0$ spin-triplet pairing.
Equal spin-triplet pairing ($m = 1)$ is produced at the TI edge if a term misaligned with the spin-quantization axis is present in the Hamiltonian. This happens for Rashba spin-orbit coupling \cite{Kane05b}:
\begin{align}
\label{eq:HR}
H_{R} = i \lambda_R \sum_{\langle i,j\rangle} c^\dagger_i({\bf s} \times {\bf \hat{d}}_{ij})_z c_i,
\end{align}
which is present when $z\rightarrow -z$ symmetry is broken. Here ${\bf \hat{d}}_{ij}$ is the unit vector along the bond between sites $j$ and $i$. $H_{\rm KM}+H_R$ is still in the non-trivial topological phase with a single Dirac cone at the edge for
 $\lambda_R < 2\sqrt{3} \lambda$.
\begin{figure}[htb]
\includegraphics[scale = 0.9]{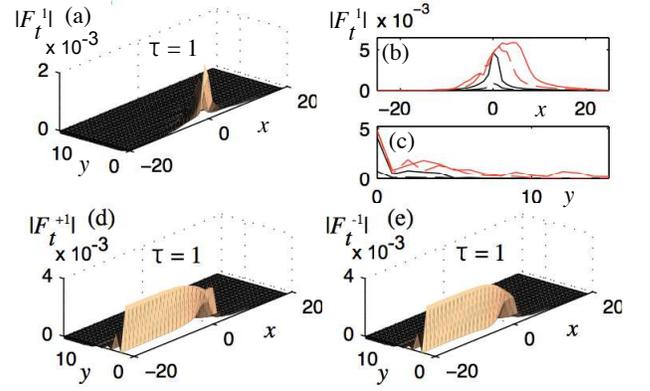}
\caption{\label{fig:Rashba} (Color online) (a) Odd-frequency pairing $|F_t^{1}(1)|$ in the SN junction in
Fig.~\ref{fig:SN} with $\lambda_R = 2\lambda$. (b) $|F_t^1|$ along the TI edge and (c) at the SN interface into the TI for $\tau = 0.5$ (dashed black), 2 (black), 8 (dashed red), and 16 (red).
(d) Spin components $|F_t^{+1}(1)|$ and (e) $|F_t^{-1}(1)|$ of $F_t^{1}(1) = (F_t^{+1}-F_t^{-1})/2$.
}
\end{figure}
We find that $F_t^0$ decreases only slightly when introducing a finite $\lambda_R$, and even for $\lambda_R = 3\lambda$, we have $F_t^0>F_t^1$. For a SN junction we find that $F_t^1$ is localized to the interface region, having a similar $x$ dependence as $F_t^0$, see Figs.~\ref{fig:Rashba}(a-c). Here we also see that $F_t^1$ extends somewhat farther into the TI, especially for larger $\tau$.
We further analyze $F_t^1$ by studying its two spin parts $F_t^{\pm 1} = \langle c_{i,\pm \sigma}(\tau)c_{i,\pm \sigma}(0)\rangle$. Surprisingly, $F_t^{\pm 1}$ is not only non-zero at the SN interface, but in the whole S region, as shown in Fig.~\ref{fig:Rashba}(d-e). In fact, we find that a non-zero $F_t^{\pm 1}$ is generated whenever $\frac{\partial \Delta}{\partial y}$ is non-zero, i.e.~for a finite order parameter gradient perpendicular to the surface of the TI. Since the low-energy density of states varies dramatically between the surface and the bulk of a TI, there will always be a strong such gradient for proximity-induced superconductivity in a TI. In the bulk of the S region $F_t^{+1} = F_t^{-1}$, and thus the criterion for $F_t^1$ to be non-zero is the same as for $F_t^0$, i.e.~a finite gradient in-surface gradient $\frac{\partial \Delta}{\partial x}$. In terms of the ${\bf d}$-vector for the $\partial_\tau F^t|_{
\tau = 0}$ odd-frequency pairing, we find that it is always real, yielding a unitary spin-triplet state.

\section{Summary}
In summary, we have shown that odd-frequency spin-triplet $s$-wave pairing is present in a TI proximity-coupled to a conventional $s$-wave spin-singlet superconductor in zero magnetic field.  The time-independent order parameter for the odd-frequency pairing is proportional to the in-plane gradient of the induced $s$-wave gap.
This disorder robust odd-frequency response is an immediate consequence of the spin-momentum locking in the TI surface state. We have explicitly demonstrated the occurrence of odd-frequency correlations not only in SN and SS' junctions at a 2D TI edge, but also when a supercurrent is applied along the edge.
In terms of experimental observables, we find no evidence of subgap states in the presence of odd-frequency pairing, due to its particular frequency dependence. The gapped LDOS could allow the detection of the odd-frequency component with local tunneling probes. Furthermore, the spin-triplet component could produce a Knight shift in NMR or muon spin-rotation measurements.
%
%
\begin{acknowledgments}
We are grateful to E.~Abrahams,  M.~Fogelstr\"om and J.~Linder for discussions. AMBS was supported by the Swedish research council (VR) and thanks LANL for hospitality where this work was initiated. Work at LANL was supported by US DoE Basic Energy Sciences and in part by the Center for Integrated Nanotechnologies, operated by LANS, LLC, for the National Nuclear Security Administration of the U.S. Department of Energy under contract DE-AC52-06NA25396.

\end{acknowledgments}


\begin{thebibliography}{24}
\bibitem[{Sup()}]{backactionnote}
\bibinfo{note}{We have found no evidence of back-action of the odd-frequency pairing on the superconducting state in the SC.}
\end{thebibliography}

\balancecolsandclearpage

%
\section{Supplementary material}
\begin{figure}[htb]
\includegraphics[scale = 0.3]{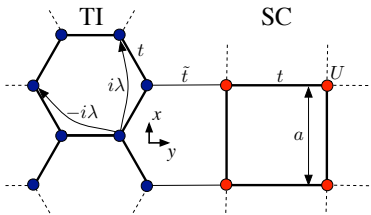}
\caption{\label{fig:SMint} (Color online) Microscopic details of the TI-SC interface with TI sites $c_i$ (dark) and SC sites $b_i$ (light).
}
\end{figure}

In this supplementary material we provide numerical data showing the absence of subgap states in the presence of odd-frequency pairing in a topological insulator (TI) induced by a gradient in the proximity-induced conventional $s$-wave superconducting state. In order to do so accurately we explicitly model the microscopic interface between a superconductor (SC) and a two-dimensional (2D) TI displayed in Fig.~\ref{fig:SMint}.

The total Hamiltonian is $H = H_{\rm KM} + H_{\rm SC} + H_{\tilde{t}}$, where $H_{\rm KM}$ defines the TI and is given by Eq.~(5) in the main text, whereas 
\begin{align}
\label{eq:HSC2}
H_{\rm SC} & = -t\sum_{\langle i,j\rangle,\sigma}b^\dagger_{i\sigma}b_{i\sigma} - U\sum_{i}b^\dagger_{i\uparrow}b_{i\uparrow}b_{i\downarrow}^\dagger b_{i\downarrow} \\
\label{eq:Htilde}
H_{\tilde{t}} & = -\tilde{t}\sum_{\langle i,j\rangle,\sigma} c^\dagger_{i \sigma} b_{i\sigma} + {\rm H.c.},
\end{align}
defines the SC and the coupling between the TI and the SC, respectively. The SC is defined on a square lattice with nearest neighbor hopping $t$ and an on-site spin-singlet $s$-wave pairing from an attractive Hubbard $U$ term. The coupling between the TI and the SC is by a tunneling element $\tilde{t}$ acting between nearest neighbors across the interface. We treat Eq.~(\ref{eq:HSC2}) self-consistently within mean-field theory by using the self-consistency condition $\Delta(i) = -U\langle b_{i\downarrow}b_{i\uparrow}\rangle$. 

\begin{figure}[h]
\includegraphics[scale = 0.9]{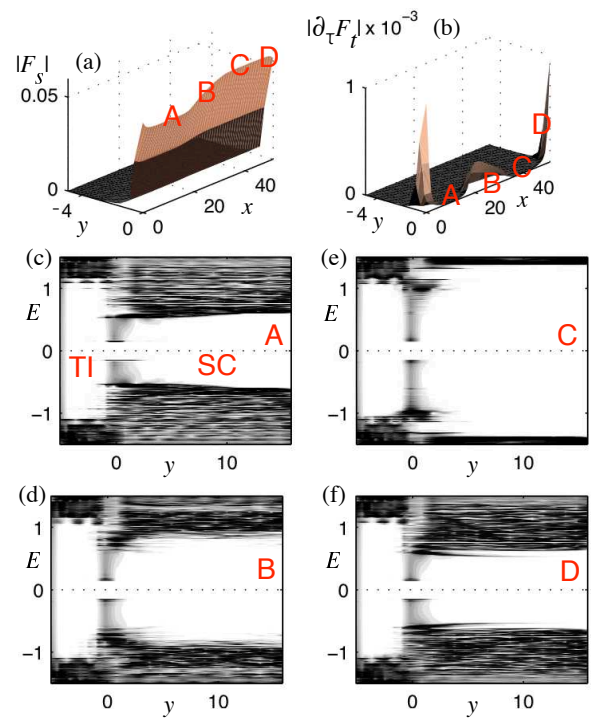}
\caption{\label{fig:LDOS} (Color online) TI-SC interface with $U = 2.5$ in the A region (18 sites), $U = 4$ in the C region (18 sites), linearly varying between these values in region C (16 sites) and an atomically sharp drop/rise in region D. (a) Magnitude of even-frequency pairing $F_s$ and (b) odd-frequency pairing $\partial_\tau F_t^0|_{\tau=0}$ in the TI. (c)-(f) LDOS plots with grayscale limits 1 (black) and 0 (white) states/site/energy in regions A-D. Dotted line mark the Fermi level. Here $\mu = 0$, $\lambda = 0.3$, and $\tilde{t} = 0.9$.
}
\end{figure}

In Fig.~\ref{fig:LDOS} we show a typical interface when the pairing potential $U$ is set to vary along the interface, i.e.~in the $x$-direction, in order to produce a gradient along the TI edge, but is constant along the $y$-direction. Region A has a constant $U = 2.5$ giving $\Delta = 0.60$ in the bulk of the SC and region C has a similarly constant $U = 4$ giving $\Delta = 1.38$. In the intervening region B there is a linear rise of $U$ between these two values. Finally, the interface is put on a cylinder, making a sharp S$(U=2.5)$-S$(U=4)$ interface at D. In Fig.~\ref{fig:LDOS}(a) we clearly see how the magnitude of the induced $s$-wave pairing $F_U$ in the TI reflects the change in $U$ along the interface. Figure~\ref{fig:LDOS}(b) displays the magnitude of the odd-frequency pairing order parameter $\partial_\tau F_t^0|_{\tau=0}$, which is only non-zero in the B and D regions. The odd-frequency pairing leaking into the SC is at least an order of magnitude smaller and we can not deduce any physical consequences in the SC from this back action.
Figures~\ref{fig:LDOS}(c)-(f) shows the corresponding local density of states (LDOS) in the A-D regions. Starting from the far left, these LDOS plots show the unperturbed left TI edge, with a constant DOS due to the one-dimensional surface Dirac cone. When we see the constant TI bulk gap of 1 before the right TI edge which is gapped by the induced superconductivity from the SC. The gap in the bulk of the superconductor equals $\Delta$ which varies significantly in $x$, but the induced gap in the TI surface state varies much less. Most notably, we see no evidence of any subgap states in the C and D region where odd-frequency pairing is present. In fact, the right TI edge in the C and D regions looks remarkably similar to a simple interpolation between the edge states in regions A and B.
\end{document}